\documentclass[aps, twocolumn]{revtex4}

\usepackage{float}
\usepackage{dcolumn}
\usepackage{amsmath}
\input{epsf}
\input{epsfx}

\ifx\pdftexversion\undefined
  \usepackage{graphicx}
\else
  \usepackage[pdftex]{graphicx}
\fi
\usepackage{epsfig}
\usepackage{ulem}

\begin{document}
\title{Fast Electrical Control of a Quantum Dot Strongly Coupled to a Nano-resonator}


\author{Andrei Faraon$^{1}$, Arka Majumdar$^{1}$, Hyochul Kim$^{2}$, Pierre Petroff$^{3}$ \& Jelena Vu\v{c}kovi\'{c}$^{*1}$}

\affiliation{$^{1}$ E. L. Ginzton Laboratory, Stanford University, Stanford CA 94305\\
$^{2}$Department of Physics, University of California, Santa Barbara, CA, 93106\\
$^{3}$Department of Electrical and Computer Engineering, University of California, Santa Barbara, CA 93106
}

%

\begin{abstract}

The resonance frequency of an InAs quantum dot strongly coupled to a GaAs photonic crystal cavity was electrically controlled via quantum confined Stark effect. Stark shifts up to 0.3meV were achieved using a lateral Schottky electrode that created a local depletion region at the location of the quantum dot. We report switching of a probe laser coherently coupled to the cavity up to speeds as high as 150MHz, limited by the RC constant of the transmission line. The coupling rate {\it g} and the magnitude of the Stark shift with electric field were investigated while coherently probing the system.

\end{abstract}

\maketitle

Photonic crystals(PCs) are one of the most promising platforms for nanophotonic networks\cite{06NodaPCreview, 2008.Faraon.DITWaveguideCoupledCav,NodaNPhotReview} to be used in information processing. Quantum dots(QDs) coupled to PC optical modes enable efficient control of light in these devices. We have already shown that the transmission function of photonic crystal cavities integrated in basic photonic crystal networks can be controlled using coupled quantum dots\cite{2008.Faraon.DITWaveguideCoupledCav,NatureRef, EdoDIT}. In addition, we have shown that the QD wavelength could be tuned to cavity resonance using local temperature tuning at speeds up to 100kHz\cite{AndreiElTtune}. However, in order to achieve higher speeds of operation as required for optical information processing devices, the QD should be controlled either optically\cite{2008.Fushman.ControlPhase,2009.Englund.ModulatorPIN} or electrically. Here we demonstrate fast electrical control via quantum confined Stark effect (QCSE) of the resonance frequency of an InAs quantum dot strongly coupled to a photonic crystal cavity\cite{Yoshie04,SCImamogluNature,AndreiTtune}. The demonstration of this type of device, that operates at the fundamental limit of light-matter interaction, adds an essential component to the toolbox of quantum optical technologies.

\begin{figure}[htbp]
    \includegraphics[width=3.5in]{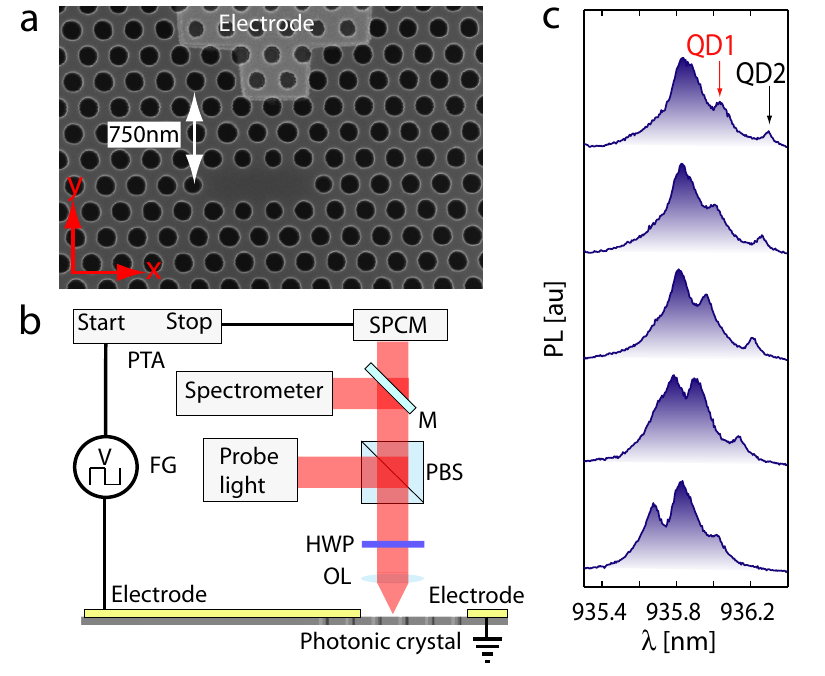}
        \caption{(a) Scanning electron microscope image of the photonic crystal cavity and the metallic electrode placed within 1$\mu m$ from the center of the cavity. (b) Schematic representation of the experimental setup (not drawn to scale). A cross-polarized confocal microscope setup composed of a polarizing beam splitter(PBS), half wave plate (HWP) and objective lens (OL) was used for photoluminescence and reflectivity measurements. The voltage on the chip was controlled using a function generator (FG) and the time domain measurements were performed using a picosecond time analyzer(PTA).(c) Photoluminescence spectra taken for different cavity/QD detunings by increasing the temperature of the sample. The avoided crossing of the polaritons indicates that QD1 is strongly coupled.}
    \label{fig:setup}
\end{figure}

The device consists of an InAs quantum dot coupled to a linear three hole defect photonic crystal cavity \cite{NodaL3} fabricated in a 160nm thick GaAs membrane (Fig.\ref{fig:setup}(a)). The electrical control was achieved by applying a lateral electric field across the quantum dot and thus shifting its resonant frequency via QCSE\cite{Hoegele04,2008.Finley.ElectricalControlSC}. The field was created in the depletion layer of a Schottky contact (20nm Cr/25nm Au on GaAs) deposited in the vicinity of the quantum dot\cite{2007.Gerardot.LateralQDControl}. A scanning electron microscope of the photonic crystal resonator integrated with the laterally positioned electrode is shown in Fig.\ref{fig:setup}(a). Another Schottky contact, located on the surface of the chip a few hundred microns away from the photonic crystal, was used to set the ground potential.

\begin{figure}[htbp]
    \includegraphics[width=3.5in]{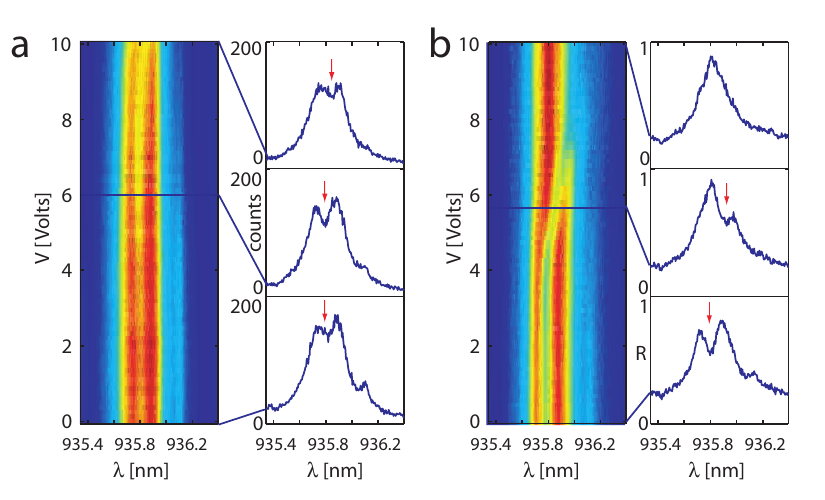}
        \caption{(a) Photoluminescence spectra as a function of  increasing control voltage ($V$) from 0V to 10V. At 0V the quantum dot was tuned on resonance with the cavity (T=48K). The PL intensity decreases and the QD red shifts for $V > 8V$. The shift is much smaller than the one observed in broadband reflectivity because of the screening induced by free carriers. (b)Broadband reflectivity spectra while changing $V$ from 0V to 10V (T=48K). The QD red shifts for $V>4V$.}
    \label{fig:PL_Ref}
\end{figure}

One challenge in designing the device is the small extent of the depletion layer in the vicinity of the Schottky contact. For typical undoped GaAs grown by molecular beam epitaxy, there is still a dopant concentration of $\sim 10^{16}/cm^{3}$ that limits the size of the depletion layer to a few microns for a 10V bias voltage. This requires the contact to be brought into a proximity of a few microns from the quantum dot embedded in the photonic crystal cavity. Since metals introduce high optical losses, the device was designed such that the metal electrode, located within $\sim 1\mu m$ from the center of the resonator, had a minimum overlap with the optical mode. The fundamental mode of the resonator extends mainly in a direction that makes an angle of $\sim 30^{o}$ with the cavity axis ($x$) and has a small extent in the $y$ direction\cite{AndreiWgCoupling}. To minimize the optical loss, the electrode was brought in the proximity of the resonator along the $y$ direction and no significant degradation of the quality factor was observed. On the same chip, we measured electrically controlled resonators with quality factors as high as 17000, similar to cavities without the metal electrode. The cavity studied in this letter had a lower quality factor ($Q\sim4000$) because it was integrated with a grating structure that allows efficient resonant in/out coupling from the resonator, as discussed in \cite{2009.Englund.CoherentExcitation}.

The photonic crystal was fabricated in a GaAs membrane as described in Ref.\cite{NatureRef} and the metal contacts were deposited by thermal evaporation. The measurements were performed at cryogenic temperatures using a cross-polarized optical setup as shown in Fig.\ref{fig:setup}(b). First, a photoluminescence (PL) measurement was performed to identify a strongly coupled QD. The signature of strong coupling is the vacuum Rabi splitting, observed (Fig.\ref{fig:setup}(c)) as an avoided crossing of the eigenstates of the system when the quantum dot is tuned into resonance with the cavity \cite{Yoshie04}. From the PL spectra one could identify two quantum dots with frequencies close to the cavity resonance, labeled as QD1 and QD2 in (Fig.\ref{fig:PL_Ref}(a)). Only QD1 showed the avoided crossing, thus indicating strong coupling. All the measurements reported in this letter were done using QD1, but the signature of QD2 was still visible in some of the data sets. For clarity, QD1 was marked with a red arrow in some of the figures. The experimental data indicated a cavity quality factor $Q\sim4000$, corresponding to a field decay rate $\kappa/2\pi\sim 40GHz$), and a quantum dot cavity coupling rate $g/2\pi \sim 20GHz$. Since $g \geq\kappa/2$ and $g >> \gamma$, ($\gamma/2\pi$ on the order of $0.1$GHz), the system operated on the onset of the strong coupling regime.

The vacuum Rabi splitting was also observed in the transmission function of the resonator, as measured using a cross-polarized reflectivity measurement(Fig.\ref{fig:PL_Ref}(b))\cite{NatureRef}. Two types of resonant probing were used in this experiment. In one case, a continuous wave (CW) laser beam was scanned through the cavity resonance and the output was monitored with a photodetector. This measurement is referred as ``CW reflectivity". In the second case, a broadband light source was coupled into the resonator and the entire reflectivity spectrum was monitored on a spectrometer. This measurement is referred as ``broadband reflectivity". While CW reflectivity gave very precise spectral information of the system due to the narrow linewidth of the laser ($\sim$300KHz), it was relatively slow. The broadband reflectivity gave all the spectral information at once but it was limited by the resolution of the spectrometer ($\sim$0.025nm).

With the quantum dot and the cavity brought into resonance (temperature set to $T=48K$), the effect of the electric field was first studied in PL by changing the bias voltage from 0V to 10V. As the bias approached $\sim 10V$, the total PL intensity decreased and the quantum dot showed a red shift of only $\sim 0.03nm$ $(0.04 meV)$ as shown in Fig.\ref{fig:PL_Ref}(a). The shift in the QD resonance was due to the QCSE, and the reduction in the PL intensity was caused by the carriers being swept away before recombining in the QD. The Stark shift and the PL reduction were only observed when using low powers of the excitation laser (tuned at 875nm). By increasing the intensity of the laser, more carriers were excited in the cavity and thus screened the electric field.

\begin{figure}[htbp]
    \includegraphics[width=3.5in]{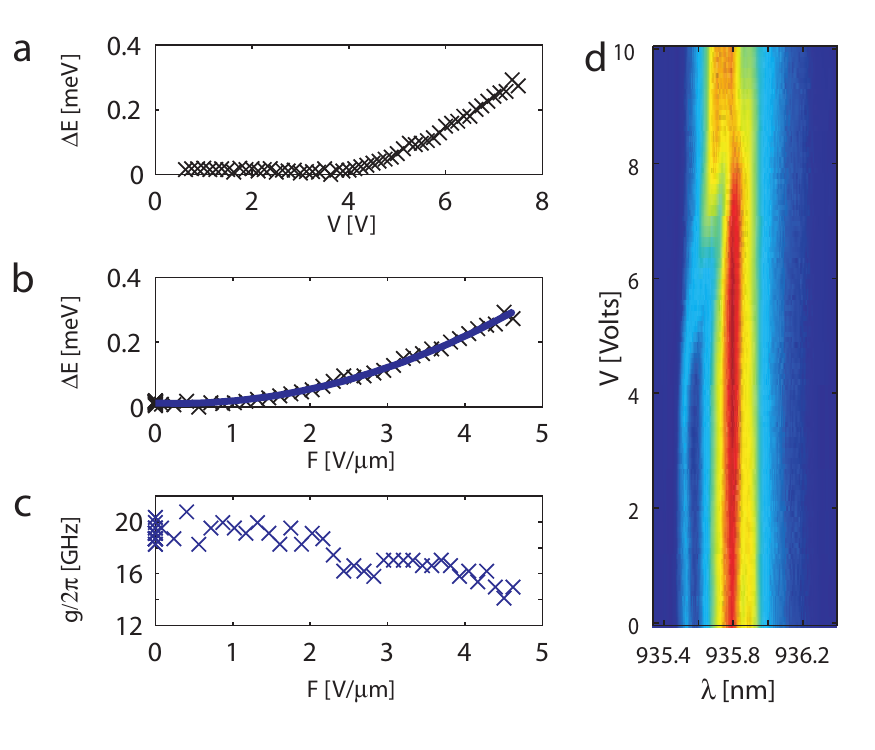}
        \caption{{\bf Change in QD frequency and coupling rate $g$ with bias voltage} (a) Stark shift of the quantum dot with applied bias. (b) Experimental data and fit indicating the quadratic dependence of the quantum dot shift with electric field. (c) Dependence of the cavity/QD coupling $g$ with applied voltage, as inferred from the fit to experimental data (d) Broadband reflectivity taken at T=46K such that the QD was resonant with the cavity at high electric fields. As the voltage approaches 8V, the signature of the quantum dot in the spectrum vanished, most probably due to the loss in the quantum dot confinement due to high electric field.}
    \label{fig:stark_shift}
\end{figure}

To test the effect of the electric field under resonant probing, the system was first measured using broadband reflectivity. A superluminescent diode with broad emission around 935nm was used as the light source, thus minimizing free carrier generation. As shown in Fig.\ref{fig:PL_Ref}(b), the effect of the bias voltage on the QD wavelength was more pronounced than in the PL measurement. The Stark shift could be observed for bias voltages larger than $V\sim 4V$, and for voltages exceeding $V\sim7V$ the quantum dot was completely detuned from the cavity. As seen from Fig.\ref{fig:PL_Ref}(c), by applying the electric field the transmission at the cavity resonance is switched from a local minimum to a local maximum. 

The dependence of the quantum dot Stark shift with the applied bias voltage was extracted from the spectra\cite{NatureRef,2008.Fushman.ControlPhase} in Fig.\ref{fig:PL_Ref}(b), and is shown in Fig.\ref{fig:stark_shift}(a).  The Stark shift is only observed for voltages larger than $V \sim 4V$, which corresponds to the depletion layer extending to the location of the quantum dot. The magnitude of the electric field in the center of the cavity was inferred by modeling the Schottky contact. The size of the depletion layer ($x_{d}(V)$) and the electric field in the cavity ($F(V)$) are given by $x_{d}=\sqrt{2\epsilon_{GaAs}(\phi-V)/(e N_{d})}$ and $F=-e N_{d} (x_{d}-\Delta x)H(\Delta x-x)/(\epsilon_{0}\epsilon_{GaAs})$. Here, $\Delta x = 750nm$ is the distance between the electrode and the center of the cavity, $N_{d}=9\times 10^{15}/cm^{3}$ is the doping concentration, $\phi=0.36V$ is the potential barrier of the Schottky contact, $e$ is the electron charge, $\epsilon_{GaAs}=12.9$ is the dielectric constant of GaAs at low temperatures, and $H(x)$ is the unit-step function. The effect of the surface states were not considered when estimating the electric field. The dependence of the energy shift with electric field is shown in Fig.\ref{fig:stark_shift}(b). The shift was quadratic in electric field, since the perturbation of the energy levels due to electric field is a second order effect.  The data was fit using\cite{2007.Gerardot.LateralQDControl} $\Delta E=\mu F-\alpha F^{2}$ with $\alpha=-0.015 meV \mu m^{2}/V^{2}=-2.4 \times 10^{-36} J m^{2}/V^{2}$ and $\mu=-0.009 meV \mu m/V =-1.4 \times 10^{-30} J m/V$.

The confining potential of the quantum dot could be perturbed by the influence of the electric field. For the data set shown in Fig.\ref{fig:PL_Ref}(b) (taken at $T=48K$) the QD became completely off resonant with the cavity for $V>7V$ so the reflectivity spectrum at high electric field yielded little information about the quantum dot. For a better investigation of the QD behavior at large electric fields, another data set was taken at $T=46K$ (Fig.\ref{fig:stark_shift}(d)) such that the QD was resonant with the cavity for $V>7V$. Under bias voltage, the electron and hole wavefunctions were deformed and pulled in opposite directions thus reducing their overlap. This resulted in a reduction of the cavity/QD coupling $g$. The fits to the data showed that $g/2\pi$ decreased from $\sim 20GHz$ to $\sim 15GHz$ when a bias of $\sim 7V$ was applied. For $V>8V$, the influence of the electric field was strong enough to completely erase the signature of the quantum dot from the broadband reflectivity spectrum. This could be either due to reduced $g$, or high tunneling rate of the electron-hole pairs out of the quantum dot placed in the electric field.

\begin{figure}[htbp]
    \includegraphics[width=3.5in]{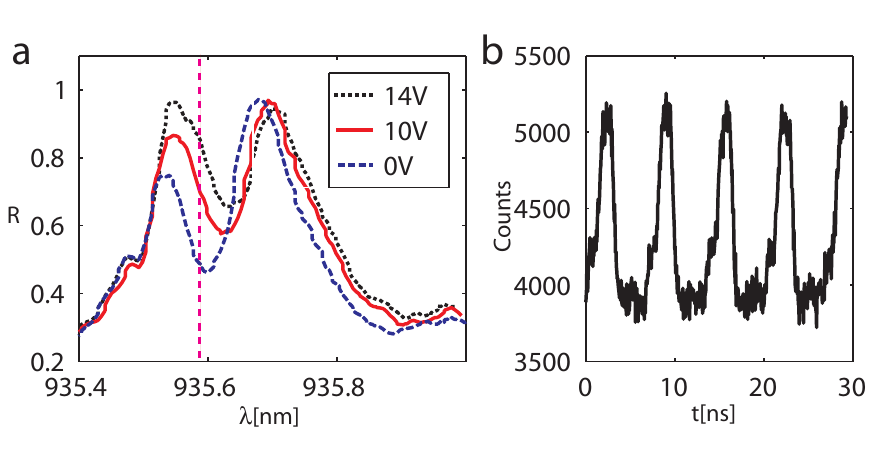}
        \caption{{\bf Fast electrical switching of a continuous wave laser beam} (a) CW reflectivity for bias voltages of 0V, 10V and 14V. During the time-domain switching experiment the laser was set at the wavelength marked by the vertical dashed line and a 0-10V signal was applied. (b) Switching of the coupled laser at the electrical QD driving frequency of 150MHz.}
    \label{fig:switching}
\end{figure}

The CW reflectivity spectra for different values of the bias voltage are shown in Fig.\ref{fig:switching}(a). The Stark shift was still present but its magnitude was smaller compared to the broadband reflectivity measurement ($0.04meV$ for $V \sim 10V$), most probably due to carriers that screen the electric field. Unlike the broadband source, the CW laser coherently created excitons in the quantum dot. Due to the bias voltage, these electron-hole pairs could tunnel out of the quantum dot and become free carriers that screened the electric field \cite{1998.Heller.PRB.ElectricFieldEffectsOnExcitons}. Alternatively, since more than one quantum dot was present in the cavity, the screening could also be caused by excitons created in the neighboring off resonant quantum dots. It has already been shown that these excitons could be created through the off resonant energy transfer between the photonic crystal resonator and the coupled quantum dots\cite{2009.Englund.CoherentExcitation,SCImamogluNature}. The screening of the electric field became more pronounced with increasing laser intensity, thus affecting the device performance. More experiments need to be performed with other devices to determine if this screening is a general property of devices based on a strongly coupled QDs, or it is particular to this system. The effect of the QD on the transmission function of the resonator can be observed for coupled probe powers as high as tens on nW (as previously shown in\cite{NatureRef}), but manipulation of the QD via QCSE at those probe power levels still needs to be demonstrated and may be limited by electric field screening. The CW reflectivity spectra in Fig.\ref{fig:switching}(a) indicated that an on/off switching ratio of $\sim 1.5:1$ was achievable with this system when driven between 0V and 10V (on/off ratio of 2:1 achievable for 0V to 14V driving). Although on:off ratios of 100:1 are theoretically expected with this system, the experimentally observed on/off ratio was limited by the properties of the quantum dot, especially operation at the onset of the strong coupling regime, decoherence\cite{2009.Englund.CoherentExcitation} and QD blinking\cite{2008.Faraon.Blockade}.

The time domain measurement was performed by setting the probe laser at the QD frequency (marked by the vertical dashed line in Fig.\ref{fig:switching}(a)) and by controlling the voltage using a function generator.  The modulated output was monitored using a single photon counting module (SPCM) and a dual channel picosecond time analyzer(PTA) synchronized to the function generator. To minimize the amount of screening due to coherently excited carriers, the probe laser power was set to $\sim 10pW$. The switching behavior at 150MHz is shown in Fig.\ref{fig:switching}(b), with an on/off ratio of $\sim 1.3:1$. This is smaller than the expected $\sim 1.5:1$ because the 3dB cutoff in the transmission line at 100MHz. An on/off ratio of 1.45:1 was observed when driving the system at 80MHz, close to the value expected from the DC measurement. 

The performance of the proof of concept device reported in this letter is limited by the experimental setup and the non-ideality of the strongly coupled system. All-optical measurements on similar devices showed that speeds up to 10GHz could be achieved with this type of system\cite{2009.Englund.CoherentExcitation}. With improved engineering, similar speeds should be achievable in electrical operation. Theoretically, when operating with $g,\kappa >> \gamma$ (i.e. strong coupling regime or high Purcell factor regime) as is the case for quantum dots in photonic crystals,  the maximum bandwidth is limited to $min(g/\pi,\kappa/\pi)$ in the strong coupling regime and $g^{2}/(\pi \kappa)$ in the weak coupling regime. Regarding the energy required to shift the QD, it is fundamentally limited by the energy density of the electric field required to shift the quantum dot inside the active volume. Considering an active volume the size of the resonator ($V_{a} \sim 1\mu m \times 1\mu m \times 200nm$), and an electric field $F \sim 5 \times 10^{4} V/cm$, this translates into a switching energy of $\sim 1fJ$, much lower than state of the art devices\cite{2008.Xu.HP.MicronScaleSiMicroring,2008.Liu.NatPhot.50fJModulator,2007.XuLipson.12GbitModulator,2004.Paniccia.Nature.HighSpeedSiliconOptMod,2009.Chen.PicoJouleOpticalSwitching}. Confining the electric field over such a small volume is not trivial, but suitable technological solutions may be found in the future.

In conclusion, we demonstrated dynamic control of QCSE in a QD strongly coupled to a photonic crystal cavity. We reported electro-optic switching up to 150MHz with an on:off ration of 1.3:1, and discussed the prospects for improving the device performance. This type of device can be integrated in a on-chip optical or quantum network\cite{2008.Faraon.DITWaveguideCoupledCav} and will be an essential building block for future optoelectronic devices for classical and quantum information processing devices operating at ultra-low energies, where fine and fast tuning of the quantum dot resonance is required.

{\bf Financial support} was provided by the Presidential Early Career Award, DARPA Young Faculty Award and Army Research Office. Part of the work was performed at the Stanford Nanofabrication Facility of NNIN supported by the National Science Foundation. A.M. was also supported by the Stanford Graduate Fellowship (Texas Instruments fellowship). The authors thank Prof. David Miller for useful discussion.

{\bf Correspondence} and requests for materials should be addressed to J. Vuckovic~(email: jela@stanford.edu).

\end{document}